\documentstyle[12pt,a4,cite,graphicx,axodraw]{article}

\newcommand{\ba}{\begin{eqnarray}}
\newcommand{\ea}{\end{eqnarray}}
\newcommand{\be}{\begin{equation}}
\newcommand{\ee}{\end{equation}}
\newcommand{\dis}{\displaystyle}

\newcommand{\noi}{\noindent}
\newcommand{\nn}{\nonumber}
\newcommand{\tr}{\mbox{\rm tr}}

\newcommand{\ksls}{\not \! k}
\newcommand{\psls}{\not \! p}

\newcommand{\pslsh}{\not \! p}

\begin{document}

\begin{titlepage}
\begin{flushright}
CAFPE-123/09\\
UG-FT-253/09\\
\end{flushright}
\vspace{2cm}
\begin{center}

{\large\bf  Theory of the Hadronic Light-by-Light Contribution to Muon
g-2\footnote{Invited talk at ``Sixth International Workshop
on Chiral Dynamics'', July 6-10  2009, Bern, Switzerland.}}\\
\vfill
{\bf  Joaquim Prades}\\[0.5cm]
 CAFPE and Departamento de
 F\'{\i}sica Te\'orica y del Cosmos, Universidad de Granada, 
Campus de Fuente Nueva, E-18002 Granada, Spain.\\[0.5cm]

\end{center}
\vfill

\begin{abstract}
\noindent
I report on the theory, recent calculations
 and present status of
the hadronic light-by-light contribution to muon $g-2$.
In particular, I report on work done together
with Eduardo de Rafael and Arkady Vainshtein where 
we get   $a^{\rm HLbL} = (10.5 \pm 2.6) \times 10^{-10}$
as our present result for this quantity.
\end{abstract}
\vfill
September 2009
\end{titlepage}
\setcounter{page}{1}
\setcounter{footnote}{0}

\section{Introduction}

\begin{figure}[hbt]
\begin{center}
\includegraphics[width=4.5cm]{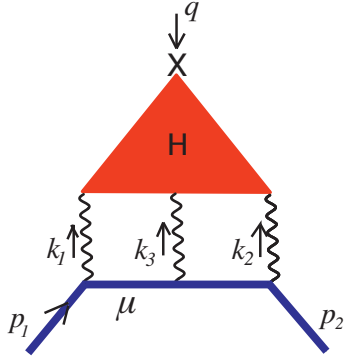}
\end{center}
\caption{Hadronic light-by-light scattering contribution.
\label{fig:1}}
\vspace*{1cm}
\end{figure}

There are six possible momenta configurations
contributing to the  hadronic light-by-light to muon g-2,
one  of them  is depicted
in Fig. \ref{fig:1} and described by the  vertex function 
\ba
\label{Mlbl}
\dis{\Gamma^\mu} (p_2,p_1)
&=&  - e^6  
\int {{\rm d}^4 k_1 \over (2\pi )^4}
\int {{\rm d}^4 k_2\over (2\pi )^4}  
{\Pi^{\mu\nu\rho\sigma} (q,k_1,k_2,k_3) 
\over k_1^2\, k_2^2 \, k_3^2} \nonumber \\ &\times&  
\gamma_\nu (\not{\! p}_2+\not{\! k}_2-m )^{-1} 
\gamma_\rho (\not{\! p}_1-\not{\! k}_1-m )^{-1} \gamma_\sigma \, 
 \nonumber \\ 
\ea
where $q \to 0 $ is the momentum of the
photon that couples to the external magnetic source,
$q=p_2-p_1=-k_1-k_2-k_3$ and $m$ is the muon mass. 

 The dominant contribution to the hadronic four-point  function 
\ba
\label{four}
\Pi^{\rho\nu\alpha\beta}(q,k_1,k_3,k_2)&=& \nonumber \\
&& i^3 \int {\rm d}^4 x \int {\rm d}^4 y
\int {\rm d}^4 z \, \, {\rm e}^{i (-k_1 \cdot x + k_3 \cdot y + 
k_2 \cdot z)}
 \,   \langle 0 | T \left[  V^\mu(0) V^\nu(x) V^\rho(y) V^\sigma(z)
\right] |0\rangle \nn  \\ 
\ea
comes from the three light-quark 
$(q = u,d,s)$ components in the electromagnetic current
$V^\mu(x)=\left[ \overline q \, \widehat Q \, \gamma^\mu \, q \right](x)$
where $\widehat Q \equiv {\rm diag} (2, -1, -1)/3$ 
denotes the light-quark electric charge matrix. 
For $g-2$ we are interested in the limit $q \to 0$ where
current conservation implies
\ba
\dis{\Gamma^\mu} (p_2,p_1) =
- \frac{a^{\rm HLbL}}{4 m} \, 
\left[\gamma^\mu, \gamma^\nu \right] \, q_\nu \, .
\ea

Therefore, the muon anomaly can then be extracted as
\ba
\label{aH}
a^{\rm HLbL}\!\!\!\!&=&\!\!\!\!\frac{e^{6}}{48m}\int\!\frac{d^4 k_1}{(2\pi)^4}\!\!\int\frac{d^4 k_2}{(2\pi)^4}\frac{1}{k_1^2 k_2^2 k_3^2}
\left[\frac{\partial}{\partial q^{\mu}}
\Pi^{\lambda\nu\rho\sigma}(q,k_1, k_3, k_2)\right]_{q=0}  \nonumber\\[1mm] 
 & &\times \,\tr\left\{
        (\psls + m)[\gamma_{\mu},\gamma_{\lambda}](\psls +m) \gamma_{\nu}(\pslsh+\ksls_2-m)^{-1}\gamma_{\rho}(\pslsh -\ksls_1-m)^{-1}\gamma_{\sigma}\right\}\,. \nn \\
\ea

Here I report on 
 the results of \cite{PRV09} and \cite{BP07}.
Previous work on the hadronic light-by-light contribution
to muon $g-2$ can be found in 
\cite{HKS95,HK98,BPP95,HK02,BPP02,KN02,KNP02,DB08,NYF09,MV04}
and recent reviews are in \cite{PRA09,EdR09,JEG08,JN09}.

The hadronic four-point function 
$\Pi^{\mu\nu\rho\sigma}(q,k_1,k_3,k_2)$
is an extremely difficult object involving  many scales and  
no full first principle calculation 
of it has been reported yet --even in 
the simpler large numbers of colors $N_c$ of QCD limit.
Notice that we need that hadronic four-point function
 with momenta $k_1$, $k_2$ and $k_3$  varying from 0 to $\infty$ and
$q \to 0$.
Unfortunately, unlike the hadronic vacuum polarization,
there is neither  a direct connection of $a^{\rm HLbL}$
 to a measurable quantity. Two lattice groups have started 
exploratory calculations \cite{HB06,RAK07}  but the 
final uncertainty that these calculations can reach is 
not clear yet.

Attending to a combined  large number of colors of QCD
 $N_c$
 and chiral perturbation theory  (CHPT) counting, one can
distinguish four types of contributions \cite{EdR94}.
Notice that we use the CHPT counting  only for organization of the
contributions and refers to the lowest order term contributing
in each case. In fact,  Ref. \cite{PRV09} shows that 
there are  chiral  enhancement factors that demand 
more than   
Nambu-Goldstone bosons in the CHPT expansion in 
the light-by-light
contribution to the muon anomaly. See more comments on this 
afterwards.

\vspace*{0.2cm}
The four different types of contributions mentioned above 
are:
\begin{itemize}
\item Nambu-Goldstone boson exchanges contribution are ${\cal O}(N_c)$
and start contributing  at ${\cal O}(p^6)$ in CHPT.
\item One-meson irreducible vertex contribution and 
non-Goldstone boson exchanges contribute  also at
${\cal O}(N_c)$  
but start contributing at ${\cal O} (p^8)$ in CHPT.
\item One-loop of Goldstone bosons contribution
 are ${\cal O}(1/N_c)$ and start at ${\cal O}(p^4)$ in CHPT.
\item One-loop of non-Goldstone boson contributions 
which are ${\cal O}(1/N_c)$  but start contributing  
at ${\cal O}(p^8)$ in CHPT.
\end{itemize}
Based on the counting above there are two full calculations
\cite{HKS95,HK98,HK02}  and \cite{BPP95,BPP02}. 
There is also a detailed study of the $\pi^0$
exchange contribution \cite{KN02} putting emphasis in obtaining
analytical expressions for this part.
Recently, two new calculations of the pion exchange
using  also the organization above have been made. In Ref.
\cite{DB08}, the pion pole term exchange is evaluated  
within an effective chiral model, N$\chi$QM.
These authors  also study the box diagram
one-meson irreducible vertex contribution. 
The results are numerically very similar to the ones found in the 
literature as  can be seen in Table \ref{tab1}.
In Ref. \cite{NYF09}, the author uses a large $N_c$ model 
$\pi^0 \gamma^* \gamma^*$ form factor with the pion also 
off-shell. 
This has to be considered as a first step and more work 
has to be done in order to have the full light-by-light within this approach. 
In particular, it would be very interesting to calculate the 
contribution of one-meson irreducible vertex contribution 
within   this model.

There is also model independent short-distance QCD 
information on the relevant form factor. 
Using  operator product expansion (OPE) in QCD, 
the authors of \cite{MV04} pointed out a short-distance
 constraint of the reduced full four-point Green function
(form factor)
\ba
\langle 0 | T \left[ 
V^\nu (k_1) V^\rho (k_3) V^\sigma (-(k_1+k_2+q)) \right]
| \gamma(q) \rangle  
\ea
when $q \to 0$  and in the special momenta configuration 
 $-ks_1^2 \simeq -k_3^2 >> -(k_1+k_3)^2$ Euclidean and large.
In that kinematical region,
\ba
T \left[ V^\nu (k_1) V^\rho (k_3)\right]
\sim \frac{1}{\hat {k}^2} \, \varepsilon^{\nu\rho\alpha\beta}
\hat k_\alpha \, \left[\overline q \, {\hat Q}^2 \, \gamma_\beta 
\gamma_5 \, q \right] 
\ea
with $\hat k = (k_1-k_3)/2 \simeq k_1 \simeq - k_3$\, .
See also \cite{KPM04}. This short distance constraint
 was not  explicitly imposed in calculations 
previous to \cite{MV04}.

\section{Leading in $1/N_c$  Results}

Using effective field theory techniques, the authors of
 \cite{KNP02} shown that the 
leading large $N_c$ contribution 
to $a^{\rm HLbL}$  contains an enhanced 
$\log^2 (M_\rho/m_\pi)$  term at low energy.
Where the rho mass $M_\rho$ acts as  an 
ultraviolet scale and the pion mass $m_\pi$ 
provides the infrared  scale.   
The leading logarithm term is generated  by 
Nambu-Goldstone boson exchange
contributions and is fixed by the Wess--Zumino--Witten (WZW) 
vertex $\pi^0 \gamma \gamma$.  
\ba
a^{\rm HLbL}(\pi^0) = \left( \frac{\alpha}{\pi} \right)^3 \, N_c 
\frac{m^2 N_c}{48 \pi^2 f_\pi^2} \, \left[ \ln^2 \, \frac{M_\rho}{m_\pi}
+ {\cal O} \left( \ln \, \frac{M_\rho}{m_\pi} \right) + {\cal O} (1) \right] 
\ea

In the chiral limit, where quark
 masses are neglected, and at large $N_c$, 
the coefficient of this  double logarithm  is model  independent 
and has been calculated    and shown to be  positive in \cite{KNP02}. 
All the calculations we discuss here 
agree with these leading behaviour  and its coefficient including the 
sign.  A global sign mistake in the $\pi^0$ exchange 
in the results presented in \cite{HKS95,HK98,BPP95}
was found by \cite{KN02,KNP02} and confirmed by \cite{HK02,BPP02} 
and by others \cite{BCM02,RW02}. The subleading 
ultraviolet scale $\mu$-dependent terms \cite{KNP02},
 namely, 
$\log(\mu/m_\pi)$ and a non-logarithmic term $\kappa(\mu)$, 
 are model dependent and calculations of them are implicit in the
results presented in  \cite{HKS95,HK98,BPP95,BPP02,MV04}. 
 In particular, $\kappa(\mu)$ contains the large $N_c$ 
contributions from  one-meson irreducible vertex 
and non--Nambu-Goldstone boson exchanges.
In the next section we review the recent model calculations
of the full leading in the $1/N_c$ expansion contributions.

\subsection{Model Calculations}

The pseudo-scalar exchange  is the dominant numerical 
contribution  and was saturated in  
\cite{HKS95,HK98,BPP95,HK02,BPP02,KN02,DB08,NYF09} 
by Nambu-Goldstone boson exchange. 
This contribution is depicted in Fig. 2
with $M \, = \, \pi^0,\eta,\eta^\prime$.
The relevant four-point function  was obtained 
in terms of the off-shell  $\pi^0 \gamma^*(k_1) \gamma^*(k_3)$
form factor ${\cal F}(k_1^2, k_3^2)$ and the off-shell
$\pi^0 \gamma^*(k_2) \gamma(q=0)$ form factor
${\cal F}(k_2^2, 0)$   modulating each one of the two 
 WZW $\pi^0 \gamma \gamma$  vertex. 
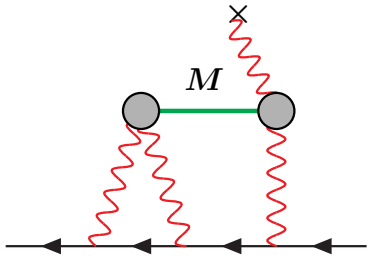
\begin{figure}
\begin{center}
\unitlength=1.6pt
\begin{picture}(80,80)
\SetScale{1.7}
\SetWidth{0.5}
\ArrowLine(20,0)(0,0)
\ArrowLine(40,0)(20,0)
\ArrowLine(60,0)(40,0)
\ArrowLine(80,0)(60,0)
\Text(55,55)[c]{\boldmath$\times$}
\SetColor{Red}
\Photon(20,0)(30,30){2}{6}
\Photon(40,0)(30,30){2}{6}
\Photon(60,0)(60,30){2}{6}
\Photon(60,30)(50,50){2}{5}
\SetColor{Green}
\SetWidth{1.}
\Line(30,30)(60,30)
\SetWidth{0.5}
\Text(47,37)[b]{\SetColor{Green}\boldmath$M$}
\SetColor{Black}
\GCirc(30,30){4}{0.7}
\GCirc(60,30){4}{0.7}
\end{picture}
\end{center}
\caption{A generic meson 
exchange contribution to the hadronic light-by-light
part of the muon $g-2$.
\label{pionexchange}}
\vspace*{1cm}
\end{figure}
 
In all cases discussed here, several short-distance QCD  constraints 
were imposed on these form-factors.
 In particular, they all have the correct QCD short-distance behaviour
\be
{\cal F}(Q^2, Q^2) \to \frac{A}{Q^2}
\hspace*{0.5cm} {\rm and} \hspace*{0.5cm} 
{\cal F}(Q^2, 0) \to \frac{B}{Q^2}
\ee
when $Q^2$ is Euclidean and large  and are  
in agreement with $\pi^0\gamma^*\gamma$ low-energy data
\footnote{See however the new 
measurement of the $\gamma \gamma^* \to \pi_0$ 
transition form factor by BaBar 
\cite{babar09} at momentum transfer
energies between 4 GeV$^2$ and 40 GeV$^2$}.
\begin{table}
\begin{center}
{\begin{tabular}{c|cc}
 References &\multicolumn{2}{c}{ $10^{10} \times a$}\\
 & $\pi^0$ only &  $\pi^0$, $\eta$ and $\eta'$\\
\hline
\cite{HKS95,HK98,HK02}  & 5.7 & 8.3 $\pm$ 0.6 \\
\cite{BPP95,BPP02} & 5.6  & 8.5 $\pm$ 1.3 \\
\cite{KN02} with $h_2=0$ & 5.8 & 8.3 $\pm$ 1.2\\
\cite{KN02} with $h_2=-10$~GeV$^2$ & 6.3 & \\
\cite{DB08} & 6.3 $\sim$ 6.7 & \\
\cite{NYF09} & 7.2 & 9.9 $\pm$ 1.6 \\
\cite{MV04} &  7.65 &11.4$\pm$1.0  
\end{tabular}}
\end{center}
\vspace*{0.2cm}
\caption{Results for the $\pi^0$,
$\eta$ and $\eta'$ exchange contributions.
\label{tab1}}
\vspace*{1cm}
\end{table}
 They  differ slightly in shape  due to the different model
assumptions (VMD, ENJL, Large $N_c$, N$\chi$QM) 
but they produce small numerical differences always compatible 
within quoted uncertainty $\sim (1.3-1.6) \times 10^{-10}$ 
--see Table \ref{tab1}.

Within the models used in \cite{HKS95,HK98,BPP95,HK02,BPP02,KN02,DB08,NYF09}, 
to get the full contribution at leading  in $1/N_c$ 
one needs to add the one-meson irreducible vertex contribution and 
the non-Goldstone boson exchanges. In particular,  below some hadronic
scale $\Lambda$,
the one-meson irreducible vertex contribution  was identified 
in \cite{BPP95,BPP02} with the ENJL quark box contribution with 
four dressed
photon legs. While to mimic the contribution of short-distance 
QCD quarks above $\Lambda$, a loop of bare massive heavy quark  
with mass $\Lambda$
and QCD vertices was used.  The results are in Table \ref{quarkL}. There, 
one can see a very nice stability region when $\Lambda$ is in the interval
[0.7, 4.0] GeV.
Similar results for  a constituent quark-box  contribution 
below $\Lambda$ were obtained in \cite{HKS95,HK98},   
though these authors didn't discuss any 
short-distance--long-distance  matching.
\begin{table}
\begin{center}
{\begin{tabular}{c|cccc}
$\Lambda$ [GeV] & 0.7 & 1.0 & 2.0 &4.0\\
\hline
\rule{0cm}{13pt} $10^{10} \times a^{\rm HLbL}$ & 2.2 &  2.0& 1.9& 2.0
\end{tabular}}
\end{center}
\caption{Sum of the short- and long-distance 
quark loop contributions  \cite{BPP95}
as a function of the matching scale $\Lambda$.
\label{quarkL}}
\vspace*{1cm}
\end{table}

In \cite{BPP95,BPP02}, non-Goldstone boson exchanges were saturated by the 
hadrons appearing in the model, i.e. the lowest scalar and pseudo-vector
hadrons. There, both states were used 
in nonet-symmetry --this symmetry is exact in the large $N_c$ limit
 of QCD.
 
Within the ENJL model, the one-meson irreducible vertex contribution  
 is related trough Ward  identities  to the scalar exchange 
which we discuss below and {\it both}  have to be included within this 
model \cite{BPP95,BPP02}.
 The result  of the scalar exchange obtained in \cite{BPP95} is 
\be
\label{scalar}
a^{\rm HLbL}(\rm Scalar) = -(0.7 \pm 0.2 ) \times 10^{-10} \, .
\ee
The scalar exchange was not included in \cite{HKS95,HK98,HK02,KN02}.
The result of the axial-vector exchanges in \cite{HKS95,HK98,HK02}
and \cite{BPP95,BPP02} can be found in Table \ref{tab3}.
\begin{table}
\begin{center}
{\begin{tabular}{c|c}
 References
 & $10^{10} \times a^{\rm HLbL}$\\
\hline
 \cite{HKS95,HK98,HK02}  & 0.17 $\pm$ 0.10  \\
\cite{BPP95,BPP02}  & 0.25 $\pm$ 0.10 
\end{tabular}}
\end{center}
\vspace*{0.2cm}
\caption{Results for the axial-vector exchange contributions from 
\cite{HKS95,HK98,HK02} and \cite{BPP95,BPP02}.\label{tab3}}
\end{table}

 Melnikov and Vainshtein used a model that saturates
the hadronic four-point function in (\ref{four}) at
leading order in the $1/N_c$ expansion by the exchange 
of the Nambu-Goldstone $\pi^0, \eta, \eta'$ and 
the lowest axial-vector $f_1$ states.
 In that model, the new OPE constraint of the reduced four-point
function found in \cite{MV04} mentioned above, 
forces the $\pi^0 \gamma^*(q) \gamma(p_3=0)$ 
vertex to be point-like rather than including a ${\cal F}(q^2,0)$ 
form factor. 
\begin{figure}
\begin{center}
\unitlength=1.6pt
\begin{picture}(80,80)
\SetScale{1.7}
\SetWidth{0.5}
\ArrowLine(20,0)(0,0)
\ArrowLine(40,0)(20,0)
\ArrowLine(60,0)(40,0)
\ArrowLine(80,0)(60,0)
\Text(54,66)[c]{\boldmath$\times$}
\SetColor{Red}
\Photon(20,0)(30,30){2}{6}
\Photon(40,0)(30,30){2}{6}
\Photon(60,0)(60,30){2}{6}
\Photon(60,30)(50,60){2}{6}
\SetColor{Blue}
\SetWidth{1.}
\Line(30,30)(60,30)
\SetWidth{0.5}
\Text(46,37)[b]{\SetColor{Green}\boldmath$\pi^0,\eta,\eta'$}
\SetColor{Black}
\GCirc(30,30){4}{0.7}
\Vertex(60,30){2}
\SetColor{Red}
\SetWidth{1.0}
\LongArrow(75,50)(62,33)
\Text(85,60)[l]{\SetColor{Green}\boldmath$Point-Like$}
\end{picture}
\end{center}
\caption{Goldstone boson exchange in the model in \cite{MV04}
contributing to the hadronic light-by-light.
\label{pionexchangeMV}}
\vspace*{1cm}
\end{figure}
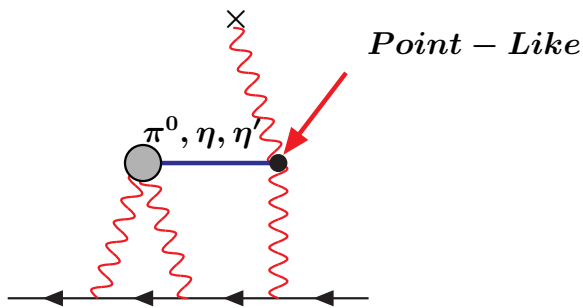
There are also OPE constraints for
other momenta regions \cite{NSV84} which are not satisfied by the model
in \cite{MV04} though the authors argued that this mismatch  makes
 only a small numerical difference of the order of $0.05 \times 10^{-10}$. 
 In fact, 
within the large $N_c$ framework, it has been shown \cite{BGL03}
that in general for other than two-point functions, to satisfy fully
the QCD short-distance properties requires
the inclusion of an infinite number of narrow states.

\section{Next-to-leading in $1/N_c$  Results}

 For the next-to-leading in $1/N_c$ contributions to
the $a^{\rm HLbL}$ there is no model independent result at present
 and is possibly the most difficult component.
 Charged pion  and kaon loops saturated this contribution
 in \cite{HKS95,HK98,BPP95,HK02,BPP02}.  To dress the photon 
interacting  with pions, a particular Hidden Gauge Symmetry (HGS) 
model was used in  \cite{HKS95,HK98,HK02} while  a full VMD was used 
in \cite{BPP95,BPP02}. 
The results obtained in these two models are
 $-(0.45\pm 0.85) \times 10^{-10}$ in \cite{HKS95}
and $-(1.9\pm0.5) \times 10^{-10}$ in \cite{BPP95} while using a point-like
vertex one gets $- 4.6 \times 10^{-10}$.

Both models (HGS and VMD)  satisfy the known constraints  
though start differing at  ${\cal O}(p^6)$ in CHPT.
 Some studies of the cut-off dependence 
of the pion loop using the full VMD model 
was done  in \cite{BPP95}  and showed  that  their  final number comes 
from fairly low energies where the model dependence should be smaller.

The authors of \cite{MV04} analyzed the model
used in \cite{HKS95,HK98} and  showed that there is a large 
cancellation between the first three terms of an expansion in powers of 
$(m_\pi/M_\rho)^2$ and  with large higher order corrections
when expanded in CHPT orders but the same applies to the $\pi^0$
exchange as can be seen from Table 6 in the first reference in \cite{BP07} 
by comparing the WZW  column with the others.
 The authors of \cite{MV04} took 
$(0\pm 1) \times 10^{-10}$ as  a guess estimate of the total NLO in $1/N_c$
contribution. This seems too simply and  certainly with 
underestimated uncertainty.

\section{Comparing Different Calculations}

The comparison of individual contributions 
in \cite{HKS95,HK98,BPP95,HK02,BPP02,KN02,DB08,NYF09,MV04}   
has to be done with care because they come from different model assumptions
to construct the full relevant four-point function. 
In fact,  the authors of \cite{DB08} have shown that their
constituent quark-box  provides the correct asymptotics
and in particular the new OPE found in \cite{MV04}. 
It has more sense to compare  results for $a^{\rm HLbL}$ either  at 
leading order or at next-to-leading order in the $1/N_c$ expansion. 

The results for the final hadronic light-by-light 
contribution to $a^{\rm HLbL}$ quoted in \cite{HKS95,HK98,BPP95,HK02,BPP02,MV04}
 are in Table \ref{tab4}.
The apparent agreement between \cite{HKS95,HK98,HK02} and \cite{BPP95,BPP02}
hides non-negligible differences which numerically almost compensate
between the quark-loop and charged pion and \cite{MV04} are in Table \ref{tab4}.
 Notice also that \cite{HKS95,HK98,HK02} didn't include the scalar exchange. 
\begin{table}
\begin{center}
{\begin{tabular}{c|c}
 Full Hadronic Light-by-Light
 & $10^{10} \times a_\mu$\\
\hline
 \cite{HKS95,HK98,HK02}  & 8.9$\pm$ 1.7  \\
\cite{BPP95,BPP02}  & 8.9 $\pm$ 3.2  \\
\cite{MV04}         & 13.6 $\pm$ 2.5 
\end{tabular}}
\end{center}
\caption{Results for the full hadronic light-by-light
contribution to $a^{\rm HLbL}$.\label{tab4}}
\end{table}
 
Comparing the results of \cite{BPP95,BPP02} and \cite{MV04},
as discussed above, we have found several differences of order
$1.5 \times 10^{-10}$  which are not related to the new short-distance
constraint used in \cite{MV04}.  The different  axial-vector mass
mixing accounts for $-1.5 \times 10^{-10}$, the absence of the
scalar exchange in \cite{MV04} accounts for $-0.7 \times 10^{-10}$
and the use of a vanishing  NLO in $1/N_c$ contribution
in \cite{MV04} accounts for $-1.9 \times 10^{-10}$. These model
dependent differences add up to $-4.1 \times 10^{-10}$ out of the
final $-5.3 \times 10^{-10}$ difference between the results
in \cite{BPP95,BPP02} and the ones in \cite{MV04} --see Table
 \ref{tab4}. 
Clearly, the new OPE constraint  used
in \cite{MV04} accounts only for a small part 
of the large numerical final difference.

\section{Conclusions and Prospects}

To give a result at present for the hadronic 
light--by--light contribution to the muon anomalous magnetic moment,  
the authors of \cite{PRV09} concluded,  from  the above considerations, 
 that it is fair to proceed as follows:

\noi
{\it Contribution to $a^{\rm HLbL}$ from $\pi^0$, $\eta$ and $\eta'$ 
exchanges}\\[1mm]
Because of the effect of the OPE constraint discussed above, 
we suggested \cite{PRV09} 
to take as central value the result of Ref.~\cite{MV04} with, however, 
 the largest error quoted in Refs.~\cite{BPP95,BPP02}:
\be
a^{\rm HLbL}(\pi\,,\eta\,,\eta')=(11.4\pm 1.3)\times 10^{-10}\,.
\ee
Recall that this central value is quite close to the one in the ENJL 
model which includes the short--distance quark-loop contribution.

\noi
{\it Contribution to $a^{\rm HLbL}$ from pseudo-vector exchanges}\\[1mm]
The analysis made in Ref. \cite{MV04} suggests that the errors 
in the first and second entries of Table \ref{tab3} are likely to be 
underestimates. Raising their $\pm 0.10$ errors  to $\pm 1$
 puts the three numbers in agreement within one sigma. We
 suggested \cite{PRV09} then as the best estimate for this
contribution at present 
\be
a^{\rm HLbL}(\rm{pseudo-vectors})=(1.5\pm 1)\times 10^{-10}\,.
\ee

\noi
{\it Contribution to $a^{\rm HLbL}$ from scalar exchanges}\\[1mm]
The ENJL--model should give a good estimate for these contributions. 
We kept \cite{PRV09}, therefore, 
the result of Ref.~\cite{BPP95,BPP02} with, 
however, a larger error which covers the effect of other
 unaccounted meson  exchanges,
\be
a^{\rm HLbL}(\rm{scalars})=-(0.7\pm 0.7)\times 10^{-10}\,.
\ee

\noi
{\it Contribution to $a^{\rm HLbL}$ from dressed charged
pion and kaon loop}\\[1mm]
Because of the instability of the results for the charged pion loop and 
unaccounted loops of other mesons, we suggested \cite{PRV09} 
using the central  value of the ENJL result  but wit a larger error:
\be
a^{\rm HLbL}(\pi{\rm -dressed~loop})=-(1.9\pm 1.9)\times 10^{-10}\,.
\ee

{}From these considerations, adding the errors in quadrature,
 as well as the small charm contribution $0.23 \times 10^{-10}$, we get
\be
a^{\rm HLbL}=(10.5\pm 2.6)\times 10^{-10}\,,
\ee
as our final estimate. 

The proposed new muon $g-2$ experiments at Fermilab \cite{HER08} 
with  $1.6 \times 10^{-10}$ accuracy goal  and at J-PARC \cite{jparc}
with  even higher  accuracy goal between $1.2 \times 10^{-10}$ and 
$0.6 \times 10^{-10}$ call for a considerable improvement in the present
calculations of $a^{\rm HLbL}$.
 The use of further theoretical and experimental constraints could
 result in reaching such accuracy soon enough. In particular,
 imposing as many as possible short-distance QCD constraints 
\cite{HKS95,HK98,BPP95,HK02,BPP02,KN02,NYF09} 
has result in a better understanding  of the numerically 
dominant $\pi^0$ exchange. 
At present, none of the  light-by-light hadronic parametrization 
satisfy fully  all short distance QCD constraints. 
In particular, this requires the  inclusion  of
 infinite number of narrow states for other 
than two-point functions  and two-point functions with soft insertions \cite{BGL03}.
A numerical dominance of certain momenta configuration
can help to minimize  the effects of short distance QCD constraints
 not satisfied, as in the model in \cite{MV04}.

More experimental information on the decays $\pi^0 \to \gamma \gamma^*$,
$\pi^0 \to \gamma^* \gamma^*$ and $\pi^0 \to e^+ e^-$ 
(with radiative  corrections included \cite{RW02,KKN06,KM09}) 
in the low- and intermediate-energy regions
(below a few GeVs)  
 can also help to confirm some of the neutral pion exchange results.
A better understanding of other smaller 
contributions but with comparable uncertainties needs both more
 theoretical work and  experimental information.
This refers in particular to  pseudo-vector exchanges. 
Experimental data on radiative decays and two-photon 
production of these and  other C-even resonances can be useful in 
that respect. 

New approaches to the pion dressed loop contribution,
 together with experimental
information on the vertex $\pi^+\pi^-\gamma^*\gamma^*$  
in the intermediate 
energy region ($0.5-1.5$) GeV would 
also be very welcome. Measurements of two-photon processes like 
$e^+e^- \to e^+e^-\pi^+\pi^-$  can be useful to give information on that 
vertex and again could reduce the model dependence.
 The two-gamma physics program low energy facilities like the experiment
KLOE-2 at DA$\Phi$NE will be very useful
 and well suited  in the processes mentioned above which
 information can help  to decrease the present model dependence 
of $a^{\rm HLbL}$.

\section*{Acknowledgments}

I would like to thank the organizers  for 
achieving such an interesting Workshop.
My thanks also to Hans Bijnens, Elisabetta Pallante, 
Eduardo de Rafael and Arkady Vainshtein for enjoyable collaborations
and Andreas Nyffeler for discussions. Work supported in part by
the European Commission (EC) RTN network Contract No. 
MRTN-CT-2006-035482 (FLAVIAnet), by MICINN, Spain  and FEDER (EC) 
Grant No. FPA2006-05294, the Spanish Consolider-Ingenio 2010 
Programme CPAN Grant No. CSD2007-00042,
and by Junta de Andaluc\'{\i}a Grant Nos. P05-FQM 437 and 
P07-FQM 03048.



\begin{thebibliography}{99}

\bibitem{PRV09}
J. Prades, E. de Rafael and A. Vainshtein in 
\textit{Lepton Dipole Moments}, B.L. Roberts and W.J. Marciano, (eds)
 (World Scientific, Singapore, 2009) 309-324,
arXiv:0901.0306.

\bibitem{BP07}
J. Bijnens and J. Prades, Mod. Phys. Lett. A 
\textbf{22} (2007) 767;
  Acta Phys.\ Polon.\  B \textbf{38} (2007) 2819.

\bibitem{HKS95}
M. Hayakawa, T. Kinoshita and  A.I. Sanda, 
Phys. Rev. Lett. \textbf{75}
(1995) 790; 
Phys. Rev. D \textbf{54} (1996) 3137.

\bibitem{HK98} M. Hayakawa and T. Kinoshita, 
Phys. Rev. D \textbf{57} (1998)465.

\bibitem{BPP95}
J. Bijnens, E. Pallante and J. Prades, 
Nucl. Phys. B \textbf{474} (1996) 379;
Phys. Rev. Lett. \textbf{75} (1995) 1447; 
Erratum-ibid. \textbf{75} (1995) 3781.

\bibitem{HK02} M. Hayakawa and T. Kinoshita, 
Phys. Rev. D  Erratum. \textbf{66} (2002) 073034. 

\bibitem{BPP02}
J. Bijnens, E. Pallante and J. Prades, 
Nucl. Phys. B \textbf{626} (2002) 410.

\bibitem{KN02}
M. Knecht and A. Nyffeler, 
Phys. Rev. D \textbf{65} (2002) 073034.

\bibitem{KNP02}
M. Knecht, A. Nyffeler, M. Perrottet and E. de Rafael,
 Phys. Rev. Lett. \textbf{88} (2002) 071802.

\bibitem{DB08}
 A.E. Dorokhov and W. Broniowski,  
Phys.\ Rev.\  D \textbf{78} (2008) 073011.

\bibitem{NYF09}
A. Nyffeler, Phys. Rev. D \textbf{79} (2009) 073012
and these proceedings.

\bibitem{MV04}
K. Melnikov and A. Vainshtein,
Phys. Rev. D \textbf{70} (2004) 113006;
A. Vainshtein,
 AIP Conf.\ Proc.\  {\bf 698} (2004) 403;
 Prog.\ Part.\ Nucl.\ Phys.\  {\bf 55} (2005) 451;
Nucl.\ Phys. B (Proc.\ Suppl.)    {\bf 162} (2006) 247;
ibid. {\bf 169} (2007) 232;
  Phys.\ Atom.\ Nucl.\  {\bf 71} (2008) 630.

\bibitem{PRA09}
J. Prades,  Nucl. Phys. B (Proc.\ Suppl.)  
\textbf{181-182} (2008) 15;
arXiv:0905.3164 (to be published in Eur. Phys. J C);
  arXiv:0907.2938.

\bibitem{EdR09} E. de Rafael,
  Nucl.\ Phys. B  (Proc.\ Suppl.)  \textbf{186} (2009) 211;
  PoS \textbf{EFT09} (2009) 050;
  D.W. Hertzog {\it et al.},  arXiv:0705.4617;
J.P. Miller, E. de Rafael and B.L. Roberts,
  Rept.\ Prog.\ Phys.\  \textbf{70} (2007) 795.

\bibitem{JEG08}
  F. Jegerlehner,  Lect.\ Notes Phys.\  \textbf{745} (2008) 9;
  Acta Phys.\ Polon.\  B \textbf{38} (2007) 3021.

\bibitem{JN09}  F. Jegerlehner and A. Nyffeler,
  Phys. Rept. \textbf{477} (2009) 1;

\bibitem{HB06}
T. Blum and S. Chowdhury, Nucl. Phys. B (Proc. Suppl.)
\textbf{189} (2009) 251;
M. Hayakawa, T. Blum, T. Izubuchi and N.Yamada,
  PoS \textbf{LAT2005} (2006) 353.

\bibitem{RAK07}
P. Rakow for QCDSF Collaboration, Talk at ``Topical Workshop
on the Muon g-2'', 25-26 October 2007, Glasgow, UK.

  \bibitem{EdR94}
E. de Rafael, Phys. Lett. B \textbf{322} (1994) 239.

\bibitem{KPM04}
M. Knecht, S. Peris, M. Perrottet and E. de Rafael,
JHEP \textbf{03} (2004) 035.

\bibitem{BCM02}
I.R. Blokland, A. Czarnecki and K. Melnikov,
  Phys.\ Rev.\ Lett.\  \textbf{88} (2002) 071803.

\bibitem{RW02}
M. Ramsey-Musolf and M.B. Wise, Phys. Rev. Lett.
\textbf{89} (2002) 041601.

\bibitem{babar09}
B. Aubert {\it et al},  [BABAR Collaboration],
  arXiv:0905.4778 [hep-ex]

\bibitem{NSV84}
V.A. Novikov, M.A. Shifman, A.I. Vainshtein, M.B. Voloshin
and V.I. Zakharov, Nucl. Phys. B \textbf{237} (1984) 525.

\bibitem{BGL03}
J. Bijnens, E. G\'amiz, E Lipartia and J. Prades, 
 JHEP \textbf{04} (2004) 055.


\bibitem{KKN06}
K. Kampf, M. Knecht and J. Novotny,
  Eur.\ Phys.\ J.\  C \textbf{46} (2006) 191.

\bibitem{KM09}
  K. Kampf and B. Moussallam,
  Phys. Rev. D \textbf{79} (2009)  076005;
  K. Kampf,     PoS \textbf{EFT09} (2009) 030.

\bibitem{HER08}
D.W. Hertzog, 
Nucl. Phys. B (Proc.\ Suppl.)  
\textbf{181-182} (2008) 5.

\bibitem{jparc}
J. Imazato,
Nucl. Phys. B (Proc. Suppl.) \textbf{129-130} (2004) 81;
``Letter of Intent: An Improved Muon $(g-2)$ Experiment at J-PARC'',
 R.M. Carey {\it et al}. (2003).

\end{thebibliography}
\end{document}